\documentclass[12pt]{article}
\pdfoutput=1

\usepackage{graphics,amssymb,float}
\usepackage{graphicx}
\usepackage{rotating}
\usepackage{dcolumn}
\usepackage{bm}
\usepackage{cite}
\usepackage{amsmath}
\usepackage{indentfirst} 
\usepackage{epstopdf}
\usepackage[usenames,dvipsnames]{xcolor}

\def\tanb{\tan\beta}

\def\eg{{\it e.g.}}

\def\gev{~{\rm GeV}}

\def\fbi{~{\rm fb}^{-1}}
\def\tev{~{\rm TeV}}
\def\bit{\begin{itemize}}
\def\eit{\end{itemize}}
\def\ben{\begin{enumerate}}
\def\een{\end{enumerate}}
\def\beq{\begin{equation}}
\def\eeq{\end{equation}}

\def\brinv{{\cal B}(H\to {\rm invisible})}

\def\neut{{\tilde{\chi}^0_1}}
\def\lsp{{\tilde{\chi}^0_1}}
\def\mneut{m_{\tilde{\chi}^0_1}}

\def\mchar{m_{\tilde{\chi}^\pm_1}}

\def\charg{{\tilde{\chi}^+_1}}

\def\lsim{\mathrel{\raise.3ex\hbox{$<$\kern-.75em\lower1ex\hbox{$\sim$}}}}
\def\gsim{\mathrel{\raise.3ex\hbox{$>$\kern-.75em\lower1ex\hbox{$\sim$}}}}
\def\ifmath#1{\relax\ifmmode #1\else $#1$\fi}

\definecolor{gris}{HTML}{A9A9A9}
\definecolor{bleuclair}{HTML}{00BFFF}
\definecolor{bleu}{HTML}{0000CD}
\definecolor{rouge}{HTML}{DC143C}
\definecolor{orange}{HTML}{FF8C00}
\definecolor{vert}{HTML}{006400}
\def\ma{M_{A^0}}
\def\tev{\text{TeV}}
\def\gev{\text{GeV}}
\def\mev{\text{MeV}}
\def\stau{\tilde{\tau}}

\def\mcha{m_{\tilde{\chi}_1^\pm}}

\def\mstaua{m_{\stau_1}}

\begin{document}
\begin{titlepage}
\begin{center}


\vspace*{2cm}
{\Large\bf LHC constraints on light neutralino dark matter\\[2mm] in the MSSM} 

\vspace*{1cm}\renewcommand{\thefootnote}{\fnsymbol{footnote}}

{\large 
Genevi\`eve~B\'elanger$^{1}$\footnote[1]{Email: belanger@lapth.cnrs.fr},
Guillaume~Drieu La Rochelle$^{2}$\footnote[2]{Email: g.drieu-la-rochelle@ipnl.in2p3.fr},
B\'eranger~Dumont$^{3}$\footnote[3]{Email: dumont@lpsc.in2p3.fr},\\[1mm]
Rohini~M.~Godbole$^{4}$\footnote[4]{Email: rohini@cts.iisc.ernet.in},
Sabine~Kraml$^{3}$\footnote[5]{Email: sabine.kraml@lpsc.in2p3.fr},
Suchita~Kulkarni$^{3}$\footnote[6]{Email: suchita.kulkarni@lpsc.in2p3.fr}
} 

\renewcommand{\thefootnote}{\arabic{footnote}}

\vspace*{1cm} 
{\normalsize \it 
$^1\,$LAPTH, Universit\'e de Savoie, CNRS, B.P.110, F-74941 Annecy-le-Vieux Cedex, France\\[2mm]
$^2\,$Universit\'e de Lyon, F-69622 Lyon, France, Universit\'e Lyon 1, CNRS/IN2P3, UMR5822 IPNL, F-69622 Villeurbanne Cedex, France\\[2mm]
$^3\,$Laboratoire de Physique Subatomique et de Cosmologie, UJF Grenoble 1,
CNRS/IN2P3, INPG, 53 Avenue des Martyrs, F-38026 Grenoble, France\\[2mm]
$^4\,$Centre for High Energy Physics, Indian Institute of Science, Bangalore 560012, India
}

\vspace{1cm}

\begin{abstract}
Light neutralino dark matter can be achieved in the Minimal Supersymmetric Standard Model 
if staus are rather light, with mass around 100~GeV.
We perform a detailed analysis of the relevant supersymmetric parameter space, including also 
the possibility of light selectons and smuons, and of light higgsino- or wino-like charginos.
In addition to the latest limits from direct and indirect detection of dark matter, ATLAS and CMS constraints 
on electroweak-inos and on sleptons are taken into account using a ``simplified models'' framework.
Measurements of the properties of the Higgs boson at 125~GeV, which constrain amongst others the 
invisible decay of the Higgs boson into a pair of neutralinos, are also implemented in the analysis. 
We show that viable neutralino dark matter can be achieved for masses as low as 15~GeV. 
In this case, light charginos close to the LEP bound are required in addition to light right-chiral staus.   
Significant deviations are observed in the couplings of the 125~GeV Higgs boson. These constitute a 
promising way to probe the light neutralino dark matter scenario in the next run of the LHC.
\end{abstract}

\end{center}

\end{titlepage}

\section{Introduction}

The discovery of a new particle~\cite{Aad:2012tfa, Chatrchyan:2012ufa,Chatrchyan:2013lba}  with properties compatible with those of a Standard Model (SM) Higgs boson has strong implications for physics beyond the SM. 
In particular, the upper limit on the branching fraction of the Higgs into  invisible and/or undetected decay modes obtained from global fits to the Higgs couplings~\cite{Belanger:2013kya,Falkowski:2013dza,Giardino:2013bma,Belanger:2013xza} is a powerful probe of models with a light dark matter candidate. 
Such dark matter candidates are motivated by hints of signals in direct detection experiments found by CoGeNT~\cite{Aalseth:2010vx,Aalseth:2011wp}, DAMA~\cite{Bernabei:2008yi}, CDMS~\cite{Agnese:2013rvf}  and CRESST~\cite{Angloher:2011uu}, 
although the interpretation of these results in terms of dark matter is challenged by negative results obtained by XENON~\cite{Aprile:2012nq,Angle:2011th,PhysRevLett.110.249901}. 
Hints of order 10~GeV dark matter might also be present in indirect dark matter searches, as discussed in, \eg, \cite{Hooper:2011ti,Hooper:2012ft,Berlin:2013dva}.
More generally, light dark matter candidates are allowed in many of the popular extensions of the SM 
and it is therefore interesting to explore this possibility irrespective of the direct detection results.  

In the Minimal Supersymmetric Standard Model, MSSM, several studies have shown that light neutralino dark matter  with mass of order 10 GeV 
can be compatible with collider data, in particular those from LEP, provided one allows for non-universality in gaugino masses~\cite{Hooper:2002nq,Bottino:2002ry,Belanger:2003wb,Bottino:2003iu,Bottino:2004qi}. 
Furthermore such light neutralinos can satisfy the recent constraints from 
B-physics observables, the muon anomalous magnetic moment, 
direct and indirect dark matter detection limits, as well as LHC limits, see~\cite{Dreiner:2009ic,Calibbi:2011ug, Arbey:2012na,Cumberbatch:2011jp,Kuflik:2010ah,Hooper:2008au,Vasquez:2010ru,
Belikov:2010yi,Boehm:2013qva,Han:2013gba,Calibbi:2013poa,Arbey:2013aba}. 

The connection between the invisible decays of the Higgs into a pair of neutralinos and the dark matter 
was explored before the discovery of the new boson, in the MSSM with non-universal gaugino masses as 
well as in the general MSSM (see for example~\cite{Belanger:2001am,AlbornozVasquez:2011aa}). The current 
precision determination of the relic density~\cite{Ade:2013zuv} and the possible constraints on the branching 
fraction of the Higgs into invisibles make further investigations of this connection very interesting. 
The precise determination of the relic density puts particularly strong constraints 
on the light dark matter.
Indeed, the mostly bino-like lightest supersymmetric particle (LSP) that is found in the MSSM  typically requires some mechanism to enhance its annihilation in order not to overclose the Universe. Possible mechanisms include $s$-channel $Z$ or Higgs exchange, or $t$-channel slepton exchange (co-annihilation with sleptons is very much limited by slepton mass bounds from LEP).
For the Higgs exchange to be efficient, one has to be close to the (very narrow) $h^0$ resonance, {\it i.e.}\ $\mneut\simeq m_{h^0}/2\simeq 63$~GeV.
The $Z$ exchange is efficient for lighter neutralinos, but requires a non-negligible higgsino component. 
Hence ${\tilde{\chi}^0_2}$ and ${\tilde{\chi}^\pm_1}$ cannot be too heavy.  
For $t$-channel slepton exchange,  the sleptons must be light, close to the LEP mass bound. 
The light neutralino scenario can therefore be further probed by searching directly for electroweak-inos and/or sleptons at the LHC~\cite{Belanger:2012jn, Boehm:2013qva}.

In this Letter, we explore the parameter space of the MSSM, searching for scenarios with light neutralinos that are consistent with all relevant collider and dark matter constraints. We extend on previous studies in two main directions: 
firstly, we take into account the current LHC limits on sleptons and electroweak-inos in an SMS (Simplified Models Spectra) approach~\cite{smodels}. 
Secondly, following \cite{Belanger:2013kya,Belanger:2013xza}, we include the fit to the properties of the 
observed 125--126~GeV Higgs boson in all production/decay channels, and we consider implications of the light neutralino dark matter scenario for this Higgs signal. 
These constraints were not taken into account in two recent studies~\cite{Boehm:2013qva,Han:2013gba}. 
Another recent paper \cite{Calibbi:2013poa} takes into account the most recent ATLAS  limits from the di-tau plus $E_T^{\rm miss}$ searches~\cite{ATLAS-CONF-2013-028}, but does not discuss implications for the Higgs signal.  

The setup of the numerical analysis is described  in Section~2. 
In Section~3, we discuss the various experimental constraints that are included in the analysis.
Our results are presented in Section~4 and conclusions are given in Section~5. 

\section{Setup of the numerical analysis}

The model that we use throughout this study is the so-called phenomenological MSSM (pMSSM) with parameters defined at the weak scale. 
The 19 free parameters of the pMSSM are the gaugino masses $M_1,\ M_2,\ M_3$, the higgsino parameter $\mu$, 
the pseudoscalar mass $M_A$,  the ratio of Higgs vev's, $\tan\beta=v_2/v_1$, 
the sfermion soft masses $M_{Q_i},M_{U_i},M_{D_i},M_{L_i},M_{R_i}$  ($i=1,\,3$ assuming degeneracy for the first two generations), and the trilinear couplings $A_{t,b,\tau}$.
In order to reduce the number of parameters to scan over, we fix a subset that is not directly relevant to our analysis to the following values: $M_3=1\ \tev$, $M_{Q_3}=750\ \gev$, $M_{U_i}=M_{D_i}=M_{Q_1}=2\ \tev$, and $A_{b}=0$.
This means that we take heavy squarks (except for stops and sbottoms) and a moderately heavy gluino. 
All the strongly interacting SUSY particles are thus above current LHC limits.
The parameters of interest are $\tanb$ and $\ma$ in the Higgs sector, the gaugino and higgsino mass parameters $M_1$, $M_2$ and $\mu$, the stop trilinear coupling $A_t$, the stau parameters $(M_{L_3}, M_{R_3}, A_\tau)$, and the slepton mass parameters $(M_{L_1}, M_{R_1})$. We allow these parameters to vary within the ranges shown in Table~\ref{tab:scanrange}.\footnote{While the resulting pattern of heavy squarks and light sleptons is not the only possible choice, it seems well motivated from GUT-inspired models in which squarks typically turn out heavier than sleptons due to RGE running. Moreover, current LHC results indicate that squarks cannot be light. For a counter-example with light sbottoms, see Ref.~\cite{Arbey:2013aba}.}
The only free parameter in the squark sector, $A_t$, is tuned in order to match the mass of the lightest Higgs boson, $h^0$, with the newly observed state at the LHC.
\begin{table}[!h]
\begin{center}
\begin{tabular}{cc|cc}
 $\tanb$ & $[5,50]$ & $M_{L_3}$ & $[70,500]$ \\
 $\ma$ & $[100,1000]$ & $M_{R_3}$ & $[70,500]$ \\
 $M_1$ & $[10,70]$ & $A_\tau$\ & $[-1000,1000]$ \\
 $M_2$ & $[100,1000]$ & $M_{L_1}$ & $[100,500]$ \\
 $\mu$ & $[100,1000]$ & $M_{R_1}$ & $[100,500]$ \\
\end{tabular}
\end{center}
\vspace*{-5mm}
\caption{Scan ranges of free parameters. All masses are in GeV.}
\label{tab:scanrange}
\end{table}

We have explored this parameter space by means of various flat random scans, some of them optimized to probe efficiently regions of interest. 
More precisely, two of our ``focused'' scans probe scenarios with light left-handed or light right-handed staus by fixing one of the stau soft mass to 500~GeV and varying the other in the $[70,150]$~GeV range. These two scans are subdivided according to the masses of the selectrons and smuons, by taking either fixed $M_{L_1}=M_{R_1}=500$~GeV or varying $M_{L_1}$ or $M_{R_1}$ within $[100,200]$~GeV. Another scan has been performed in order to probe scenarios with large stau mixing and light selectrons and smuons. In this case, $M_{L_3}$ and $M_{R_3}$ are varied within $[200,300]$~GeV and $M_{R_1}$ is tuned so that $m_{\tilde{e}_R}\in[100,200]$~GeV. 

In the following, we present the results for the combination of all our scans. 
The density of points has no particular meaning, as it is impacted by the arbitrary choice of regions of interest.
The computation of all the observables has been performed within \texttt{micrOMEGAs 
3.1}~\cite{Belanger:2013oya}. \texttt{SuSpect 2.41}~\cite{Djouadi:2002ze} has been used for the 
computation of the masses and mixing matrices for Higgs particles and superpartners, while branching 
ratios for the decays of SUSY particles have been computed with 
\texttt{CalcHEP}~\cite{Belyaev:2012qa}.

\section{Experimental constraints}

The various experimental constraints that we use in the analysis are listed in 
Table~\ref{tab:expconst}. A number of ``basic constraints'' are  imposed for a first selection. They 
include the LEP results for the direct searches for charginos and staus\footnote{Note that 
selectrons and smuons are safely above the LEP bound~\cite{leplimit} since $M_{L_1} > 100\ 
\gev$ and $M_{R_1} > 100\ \gev$.}~\cite{leplimit} and for invisible 
decays of the $Z$ boson~\cite{ALEPH:2005ab}, in addition to the OPAL limit on $e^+e^- \to \tilde{\chi}^0_{2,3} \tilde{\chi}^0_1 \to Z^{(*)}(\to q\bar{q})\tilde{\chi}^0_1$~\cite{Abbiendi:2003sc}. The anomalous magnetic moment of the muon is also 
required not to exceed the bound set by the E821 experiment~\cite{Bennett:2006fi,Hagiwara:2011af}, 
and the flavor constraints coming from $b \to s\gamma$~\cite{Misiak:2006zs,Amhis:2012bh} and from $B_s \to 
\mu^+\mu^-$~\cite{bsmumu} are taken into account. Finally, the ``basic constraints'' also require 
the 
lightest Higgs boson, $h^0$, to be within 3 GeV of the best fit mass from 
ATLAS~\cite{Aad:2013wqa} and CMS~\cite{CMS-PAS-HIG-13-005}. 
This range is completely dominated by the estimated theoretical uncertainties on the Higgs mass in 
the MSSM.

{\renewcommand{\arraystretch}{1.3}
\begin{table}[t]
\begin{center}
\begin{tabular}{c|c}
 LEP limits & $\mcha >100 \ \gev$ \\
                  & $\mstaua > 84-88 \ \gev$ (depending on $\mneut$) \\
                  & $\sigma(e^+e^- \to \tilde{\chi}^0_{2,3} \tilde{\chi}^0_1 \to Z^{(*)}(\to q\bar{q})\tilde{\chi}^0_1) \lesssim 0.05$ pb \\
 \hline
 invisible $Z$ decay & $\Gamma_{Z \to \neut\neut} < 3\ \mev$ \\
 \hline
 $\mu$ magnetic moment & $\Delta a_\mu < 4.5 \times 10^{-9}$ \\
 \hline
 flavor constraints & ${\rm BR}({b\to s\gamma}) \in [3.03,4.07] \times 10^{-4}$ \\
                             & ${\rm BR}({B_s\to\mu^+\mu^-}) \in [1.5,4.3] \times 10^{-9}$ \\
 \hline
 Higgs mass & $m_{h^0} \in [122.5,128.5]\ \gev$ \\
 \hline
 $A^0, H^0 \to \tau^+\tau^-$ & CMS results for ${\cal L} = 17\rm{\ fb}^{-1}$, $m_h^{\rm max}$ scenario \\
 \hline
 Higgs couplings & ATLAS, CMS and Tevatron global fit, see text \\
 \hline
 relic density & $\Omega h^2<0.131$ or $\Omega h^2 \in [0.107,0.131]$ \\
 \hline
 direct detection & XENON100 upper limit \\
 \hline
 indirect detection & Fermi-LAT bound on gamma rays from dSphs \\
 \hline
 $pp \to \tilde{\chi}^0_2 \tilde{\chi}^\pm_1$ & Simplified Models Spectra approach, see text \\
 $pp \to \tilde{\ell}^+\tilde{\ell}^-$ & \\
\end{tabular}
\end{center}
\vspace*{-5mm}
\caption{Experimental constraints implemented in the analysis. For details, see text.}
\label{tab:expconst}
\end{table}

In addition to the set of basic constraints, limits from searches for Higgs bosons at the LHC are taken into account. 
The heavier neutral Higgses, $A^0$ and $H^0$, are constrained by dedicated searches in the $\tau^+\tau^-$ 
channel. For these, we use the most recent limits from CMS~\cite{CMS-PAS-HIG-12-050}, given in the 
($M_{A^0}$, $\tan \beta$) plane in the $m_h^{\rm max}$ scenario, which provides a conservative lower bound in the MSSM~\cite{Carena:2013qia}.\footnote{This is particularly the case in our study because our preferred very light neutralino scenarios have a small value for $\mu$ of order 200~GeV.}
The couplings of the observed Higgs boson at around 125.5 GeV, identified with $h^0$, are constrained following the procedure of  Ref.~\cite{Belanger:2013xza}, {\it i.e.}\ making use of the information given in the 2D plane $(\mu_{\rm ggF+ttH}, \mu_{\rm VBF+VH})$ for each final state provided by the LHC experiments.\footnote{The use of 2D signal strengths has first been introduced in Ref.~\cite{Cacciapaglia:2012wb}.} These ``signal strengths ellipses'' combine ATLAS and CMS results (plus results from Tevatron) for the  four effective final states that are relevant to the MSSM: $\gamma\gamma$, $VV = WW+ZZ$, $b\bar b$, and $\tau\tau$. All the experimental results up to the LHCP 2013 conference~\cite{Belanger:2013xza} are included in the present analysis. The signal strengths are computed from a set of reduced couplings 
($C_V,\, C_t,\, C_b,\, C_\tau,\, C_g$ and $C_\gamma$)
that are computed with leading order analytic formulas, except for the couplings of the Higgs to $b$ quarks, where loop corrections are included through $\Delta m_b$~\cite{Carena:1994bv}. A given point in parameter space is considered as excluded if one of these four 2D signals strengths falls outside the 95\% confidence level (CL) experimental region.

Regarding dark matter limits, the following constraints are applied: direct detection with the 
spin-independent limit from XENON100~\cite{Aprile:2012nq} and relic density from the combined 
measurement released by Planck in Ref.~\cite{Ade:2013zuv}. 
The calculation spin-independent scattering cross section depends on nuclear parameters; 
we use $m_u / m_d=0.553$, $m_s / m_d=18.9$, $\sigma_{\pi N}=44$~MeV and $\sigma_{s}=21$~MeV. 
For the relic density, multiple ranges are given in~\cite{Ade:2013zuv}; 
we use the ``Planck+WP+BAO+highL'' best fit value of $\Omega h^2=0.1189$
assuming a theory dominated uncertainty of 10\% in order to account for unknown higher-order effects to the annihilation cross section. We will thus use $\Omega h^2<0.131$ as an upper bound or 
$0.107<\Omega h^2<0.131$ as an exact range.  
We also  consider indirect detection limits from dwarf spheroidal satellite galaxies (dSphs) released by Fermi-LAT based on measurements of the photon flux
\cite{Ackermann:2011wa}; however, given that astrophysical uncertainties are still large and that current results do not strongly constrain scenarios of interest, we do not apply them to exclude parameter points but show the values of $\sigma v$ separately.

\subsection{LHC limits on sleptons, charginos and neutralinos}

Based on the latest data at $\sqrt{s} = 8\ \tev$, the ATLAS and CMS experiments have performed a number of searches for sleptons and electroweak-inos in final states with leptons and missing transverse energy, $E_T^{\rm miss}$. These have resulted in a significant improvement over the LEP limits and therefore need to be taken into account. Direct slepton production has been considered by ATLAS~\cite{ATLAS-CONF-2013-049} and CMS~\cite{CMS-PAS-SUS-12-022} in the $\ell^+\ell^- +\, E_T^{\rm miss}$ channel;\footnote{Ref.~\cite{CMS-PAS-SUS-12-022} has been very recently updated with full luminosity at 8 TeV~\cite{CMS-PAS-SUS-13-006}. This update has not been included in the present study.}
here only limits on  selectrons and smuons are currently available.
Electroweak-ino production is usually dominated by the $pp \to \tilde{\chi}^0_2 \tilde{\chi}^\pm_1$ process, which is searched for by ATLAS~\cite{ATLAS-CONF-2013-035} and CMS~\cite{CMS-PAS-SUS-12-022} in the trilepton + $E_T^{\rm miss}$ channel. The $\tilde{\chi}^0_2$ can decay either through an on-shell or off-shell $Z$ or a slepton, while the $\tilde{\chi}^\pm_1$ can decay through an on-shell or off-shell $W^\pm$ or a slepton.

The results of these searches are often interpreted by the experimental collaborations in the form of 95\% CL upper 
limits on the cross sections times branching ratios, $\sigma\times{\rm BR}$, for various SMS topologies (for a concise overview, see\cite{Chatrchyan:2013sza,Okawa:2011xg}), and it is this SMS approach that we use to implement LHC constraints in our analysis. A typical example is $pp \to \tilde{\chi}^0_2 \tilde{\chi}^\pm_1$ followed by $\tilde{\chi}^0_2 \to Z^{(*)}\tilde{\chi}^0_1$ and $\tilde{\chi}^\pm_1 \to W^{(*)}\tilde{\chi}^0_1$ with 100\% branching fraction. In this case, 95\%~CL upper limits on $\sigma\times{\rm BR}$ are given in the $(\mneut, m_{\tilde{\chi}^0_2} = m_{\tilde{\chi}^\pm_1})$ plane. 

Each scan point in the MSSM parameter space, which survives the basic constraints as well as the Higgs and dark
matter constraints discussed above, is decomposed into its relevant SMS topologies (including the correct branching ratios) and compared against the limits given by the experiments using the {\tt SModelS} technology, which will be described elsewhere \cite{smodels}. A point is considered as excluded if one of the predicted $\sigma\times{\rm BR}$ exceeds the experimental upper limit, and allowed otherwise. This approach is conservative as it considers separately the signals coming from different SUSY particles that lead to the same final state.

In the present analysis, the SMS results used are: {\it i)} $\tilde{\ell}^\pm_L \tilde{\ell}^\mp_L \to \ell^\pm \neut \ell^\mp \neut$ and $\tilde{\ell}^\pm_R \tilde{\ell}^\mp_R \to \ell^\pm \neut \ell^\mp \neut$  from both ATLAS~\cite{ATLAS-CONF-2013-049} and CMS~\cite{CMS-PAS-SUS-12-022}, {\it ii)} $\tilde{\chi}^0_2 \tilde{\chi}^\pm_1 \to Z^{(*)}\tilde{\chi}^0_1 W^{(*)}\tilde{\chi}^0_1$  again from ATLAS~\cite{ATLAS-CONF-2013-035} and CMS~\cite{CMS-PAS-SUS-12-022}, and {\it iii)} $\tilde{\chi}^0_2 \tilde{\chi}^\pm_1 \to \tilde{\ell}_R^\pm \nu \tilde{\ell}_R^\pm \ell^\mp \to \ell^\pm \neut \nu \ell^\pm \neut \ell^\mp$ from CMS~\cite{CMS-PAS-SUS-12-022}, where $\tilde{\ell}_R$ can be a selectron, a smuon or a stau. Note that the SMS limits given by the experimental collaborations in terms of $\tilde{\chi}^0_2 \tilde{\chi}^\pm_1$ production apply for any $\tilde{\chi}^0_i \tilde{\chi}^\pm_j$ ($i=2,3,4$; $j=1,2$) combination.
Some more remarks are in order. First, for SMS results involving more than two different SUSY particles, assumptions are made on their masses ({\it e.g.}\ degeneracy of $\tilde{\chi}^\pm_1$ and $\tilde{\chi}^0_2$ or specific relations between the masses in cascade decays) that are not always realized in the parameter space we consider. We allow up to 20\% deviation from this assumption in the analysis.
Second, the results for electroweak-ino production with decay through intermediate sleptons depend on the fractions  
of selectrons, smuons and staus in the cascade decay. 
When chargino/neutralino decays into staus as well as into selectrons/smuons are relevant, we use the results 
for the ``democratic'' case from \cite{CMS-PAS-SUS-12-022} if the branching ratios into the three flavors are nearly equal (within 20\%), and those for the ``$\tau$-enriched'' case otherwise.\footnote{We also apply the democratic case if decays into selectrons/smuons are more important then those into staus, but this hardly ever occurs for the scenarios of interest.}
Moreover, the results are provided by CMS for three specific values of $x=m_{\tilde{l}}/ (\mneut + m_{\tilde{\chi}^0_2})=0.05,0.5,0.95$; we use a quadratic interpolation to obtain a limit for other $x$ values. 
However, many of the scenarios we consider have light staus and heavy selectrons and smuons, 
for which the  ``$\tau$-dominated'' case applies.
Unfortunately, this has been provided by CMS only for $x=0.5$, corresponding to $m_{\tilde{\tau}_R}= (\mneut + m_{\tilde{\chi}^0_2})/2$. To get a limit for different mass ratios, we assume that the $x$ dependence is the same as in the $\tau$-enriched case; we estimate that the associated uncertainty is about a factor of 2, and we will flag the points affected by this uncertainty in the presentation of the results. 
Note that, unfortunately, we cannot use the ATLAS $2\tau$'s + $E_T^{\rm miss}$ analysis~\cite{ATLAS-CONF-2013-028}  in this approach, as it interprets the results only as left-handed staus and only for 
$m_{\tilde{\tau}_L} = (\mneut + m_{\tilde{\chi}^0_2})/2$. 
Parameter points with left-handed staus which are likely to be constrained by this analysis ({\it i.e.}, points satisfying $m_{\tilde{\chi}^{\pm}_1} < 350$~GeV and ${\rm BR}(\tilde{\chi}^{\pm}_1 \rightarrow \tilde \nu_\tau + \tau) > 0.3$) will also be flagged.

Finally, direct production of neutralino LSP's  can only be probed through mono-photon and/or mono-jet events. 
Limits from ATLAS and CMS have been given in \cite{CMS-PAS-EXO-12-048,ATLAS-CONF-2012-147} and interpreted as limits  on spin-independent interactions of dark matter with nucleons. We do not take into account these limits since they  can only be reliably interpreted in models where heavy mediators are responsible for the neutralino interactions with quarks.  This is not the case in  the MSSM where the Higgs gives the dominant contribution to the neutralino interactions with nucleons.
 
\section{Results}

Let us now present the results of our analysis. 
Figure~1 shows the effect of the dark matter constraints. Here, 
the cyan points are all those which fulfill the ``basic constraints'' and also pass 
the limit on $A^0, H^0 \to \tau^+ \tau^-$ from CMS \cite{CMS-PAS-HIG-12-050}; 
blue points are in addition compatible at 95\%~CL with all Higgs signal strength measurements, based on the global fit of~\cite{Belanger:2013xza}; 
red (orange) points obey moreover the relic density constraint $\Omega h^2<0.131$ ($0.107<\Omega h^2<0.131$) 
and abide the direct detection limits from XENON100 on $\sigma_{\rm SI}$~\cite{Aprile:2012nq}.\footnote{To account for the lower local density when the neutralino relic density is below the measured range, the predicted $\sigma_{\rm SI}$ is rescaled by a factor $\xi = \Omega h^2/0.1189$.} 
These red/orange points also pass the LHC limits  on charginos, neutralinos and sleptons; the set of points which fulfill all constraints including those from dark matter but are excluded by LHC searches are shown in grey (underlying the red/orange points).    
We notice that typically the LHC limits reduce the density of points but do not restrict any further the 
range of masses that were allowed by the other constraints. 

The upper bound on the relic density imposes a lower limit on the neutralino mass of approximately 15~GeV while the direct detection constraint does not modify the lower limit as will be discussed below. Moreover, the relic density  constrains the parameter space and the sparticles properties especially for neutralinos with mass below $\approx  30~{\rm GeV}$. 
These are  associated with light staus and light charginos as illustrated in Fig.~\ref{fig:stau1chargino1}.
The light staus are  mostly right-handed to ensure efficient annihilation since the coupling of the bino LSP is proportional to the hypercharge which is largest for $\tilde\tau_R$. Furthermore, annihilation through stau exchange is not as efficient if staus are mixed since there is a destructive interference between the L--R contributions. The light charginos are mostly higgsino since a  small value for $\mu$ is required  to have an additional  contribution from $Z$ and/or Higgs exchange, both dependent on the LSP higgsino fraction. 

\begin{figure}[t!]\centering
\includegraphics[width=0.48\textwidth]{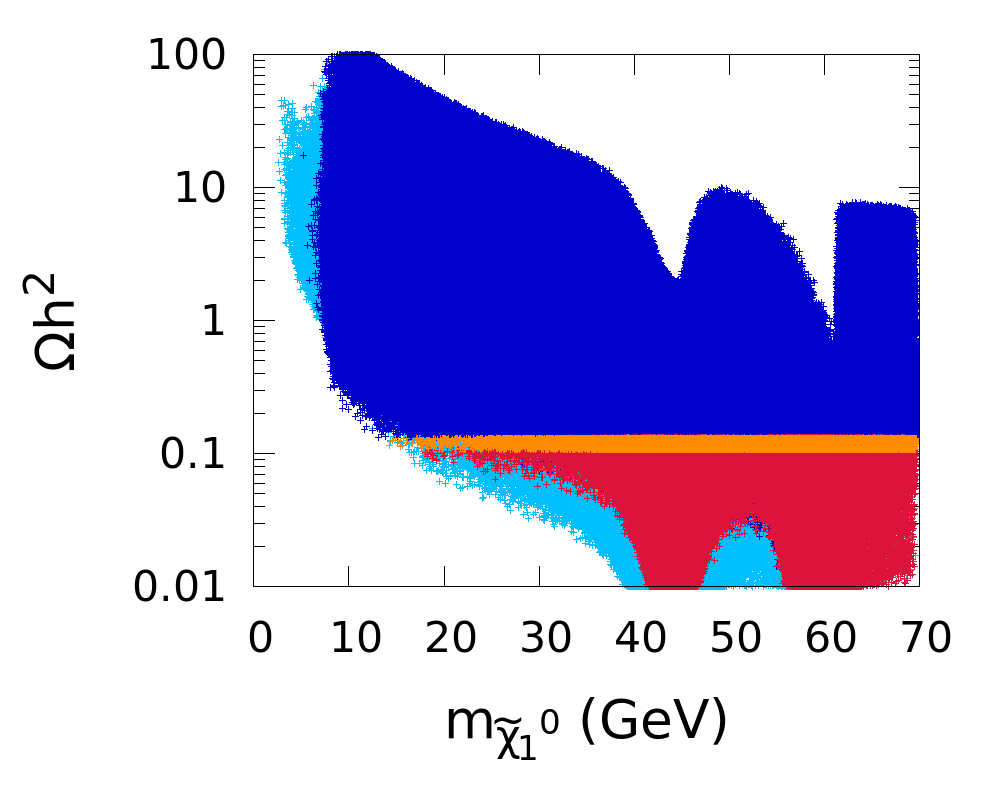}
\includegraphics[width=0.48\textwidth]{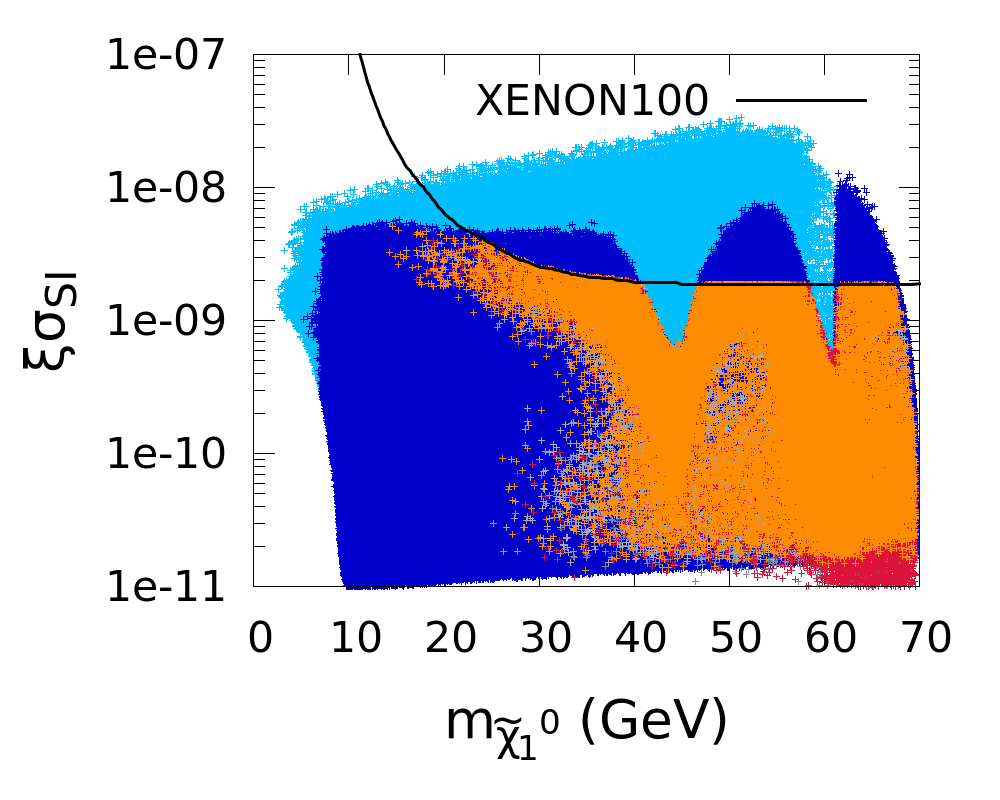}
\caption{Relic density $\Omega h^2$ (left) and rescaled spin independent scattering cross section $\xi 
\sigma_{\rm SI}$ (right) as function of the LSP mass, with $\xi = \Omega h^2/0.1189$. 
Cyan points  fulfill the ``basic constraints'' and also pass 
the limit on $A^0, H^0 \to \tau^+ \tau^-$ from CMS; 
blue points are in addition compatible at 95\%~CL with all Higgs signal strengths based on the global fit of~\cite{Belanger:2013xza}. Finally,
red (orange) points obey also the relic density constraint $\Omega h^2<0.131$ ($0.107<\Omega h^2<0.131$) 
and abide the direct detection limits from XENON100 on $\sigma_{\rm SI}$. 
\label{fig:relic} }
\end{figure}

\begin{figure}[t!]\centering
\includegraphics[width=0.48\textwidth]{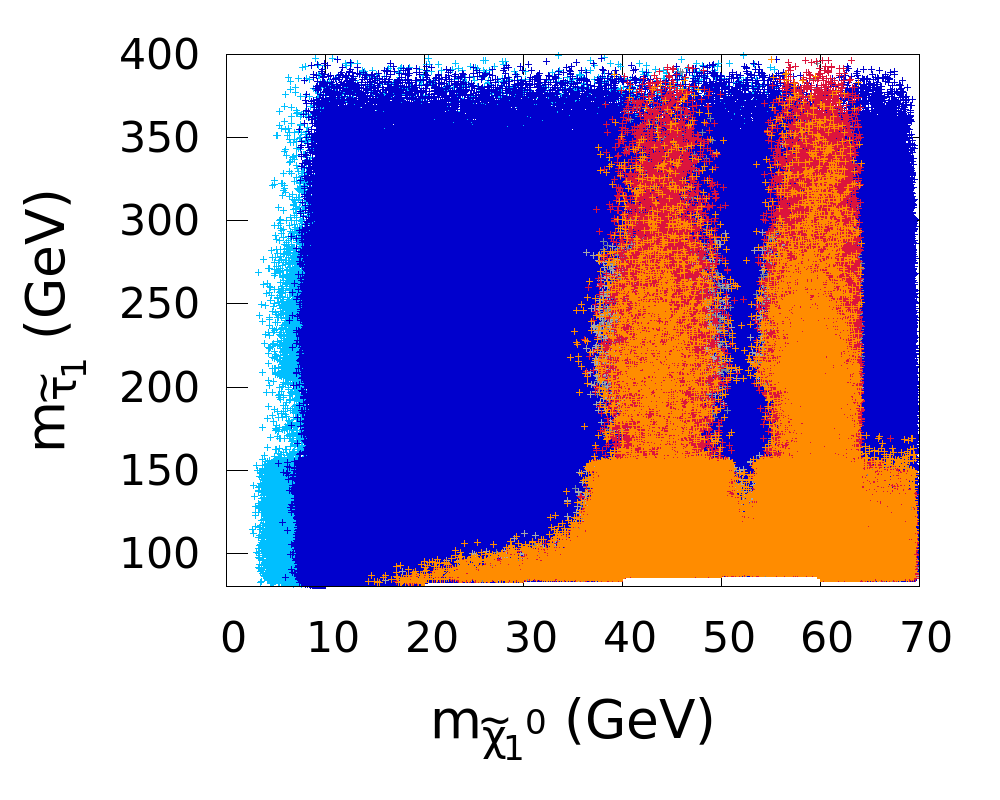}
\includegraphics[width=0.48\textwidth]{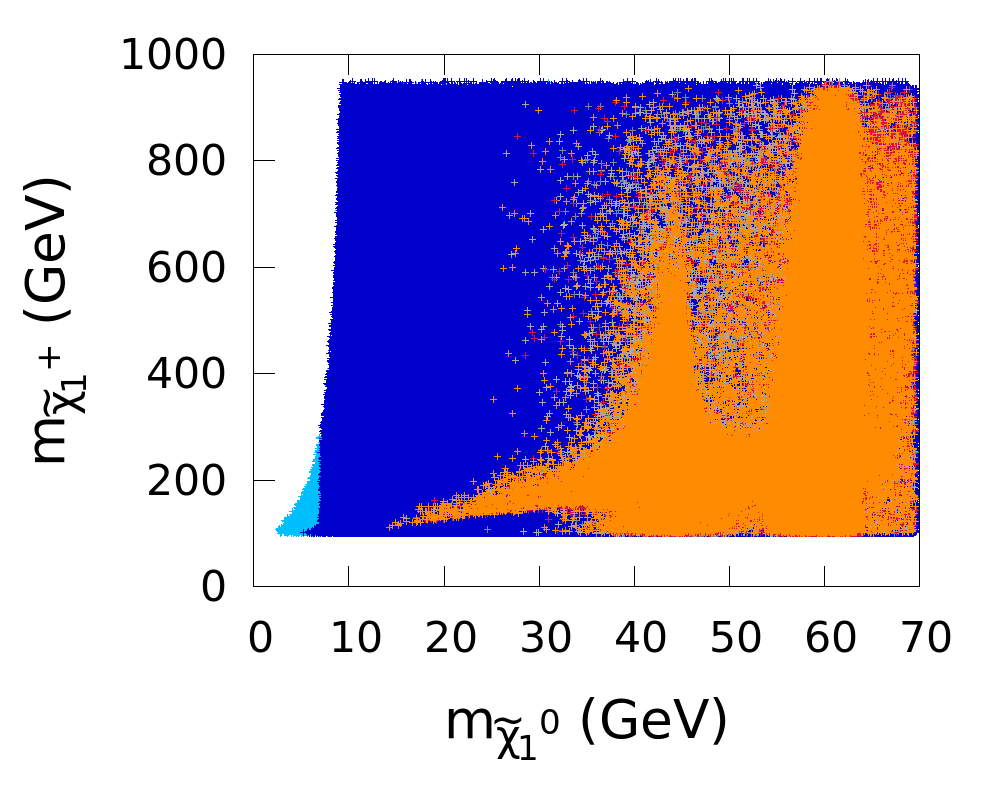}
\caption{Lighter stau mass (left) and chargino mass (right) versus $\mneut$; same color code as in Fig.~\ref{fig:relic}. 
\label{fig:stau1chargino1} }
\end{figure}

For neutralinos with masses above $\approx  30~{\rm GeV}$, the contribution of light selectrons/smuons in addition to that of the stau can bring  the relic density in the Planck range, in this case  it is not necessary to have a light chargino. These points  correspond to the scatter points with heavy charginos in Fig.~\ref{fig:stau1chargino1} (right panel).
Finally, as  the LSP mass approaches $m_Z/2$ or  $m_h/2$ the higgsino fraction can be small because of the resonance enhancement in LSP annihilation---hence the chargino can be heavy. Moreover, for $m_{\neut}\gtrsim 35$~GeV the stau contribution to the LSP annihilation is not needed, so $m_{\tilde\tau_1}$ can be large. 
Figure~\ref{fig:muM2} summarizes the allowed parameter space in the $\mchar$ versus $\mstaua$ plane (left) as well as in the $M_2$ versus $\mu$ plane (right) for different ranges of LSP masses. The $M_2$ versus $\mu$ plot illustrates the fact that when the LSP is light, $\mu$ is small, hence $\charg$ and $\tilde\chi_2^0$ are dominantly higgsino as discussed above. 
In this plot also the points for which our implementation of LHC constraints in the SMS approach has some 
significant uncertainty (from our extrapolation for the $\tau$-dominated case from \cite{CMS-PAS-SUS-12-022} 
or because the ATLAS di-tau + $E_T^{\rm miss}$ analysis~\cite{ATLAS-CONF-2013-028} is sensitive to this region in parameter space) become clearly visible. 
These points are flagged as triangles in a lighter color shade.  For $m_{\neut}<35$~GeV they concentrate 
in the region  $M_2,\mu\lesssim 320$~GeV (although a few such points have larger $\mu$). 
Most of these triangle points actually have a light $\tilde\tau_L$ and are thus likely to be excluded by the ATLAS 
result~\cite{ATLAS-CONF-2013-028}, see~\cite{Calibbi:2013poa}.
Note also that the production cross section for higgsinos is  low, so most of the points with low $\mu$ and larger 
$M_2$ are allowed. 

Another class of points that is strongly constrained by the LHC is characterized with light selectrons.   
The best limit comes from the ATLAS analysis~\cite{ATLAS-CONF-2013-049}; for LSP masses above 20 GeV, the ATLAS  searches are however insensitive to $\tilde{e}_R$ masses just above the LEP limit, more precisely in the range $m_{\tilde{e}_R} \approx 100$--120~GeV,
thus many points with light selectrons are still allowed. Furthermore, in many cases we have selectrons decaying into $\nu\tilde\chi_1^\pm$ and/or $e\tilde\chi_2^0$, thus avoiding the LHC constraint. All in all we find that 
for $\mneut>35$ GeV the whole selectron mass range considered in our scans is allowed (\textit{i.e.} either $[100,200]$ GeV or around $500$ GeV), while for $\mneut<35$ GeV, the ATLAS search 
imposes $m_{\tilde{e}_R} \approx 100$--120~GeV or ${\tilde{e}_R}$ being heavy, with the range $m_{\tilde{e}_R} \approx 120$--200~GeV being excluded. (Since we are mostly interested in how low the $\tilde\chi_1^0$ can go, we did not attempt to derive the upper end of the exclusion range for selectrons; note however that the bounds given by ATLAS vary between 230 and 450 GeV depending on the scenario.) 

\begin{figure}[t!]\centering
\includegraphics[width=0.48\textwidth]{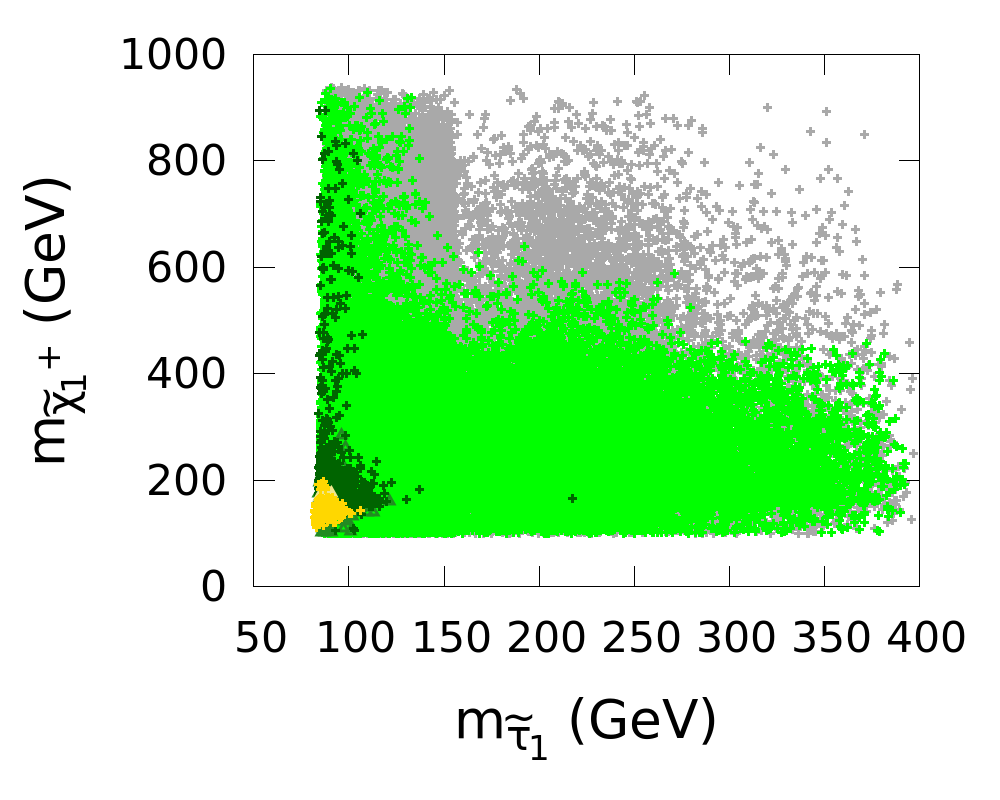}
\includegraphics[width=0.48\textwidth]{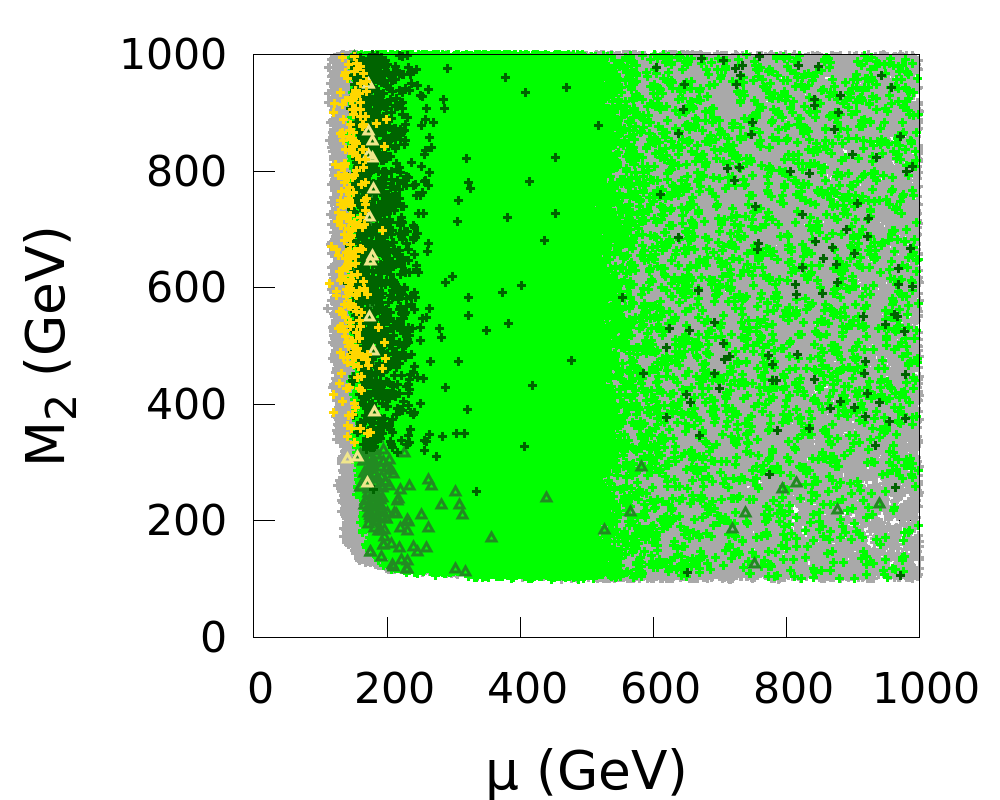}
\caption{Points passing all constraints, including $\Omega h^2<0.131$, XENON100 limits and SMS 
limits from the LHC SUSY searches: on the left in the chargino versus stau mass plane, 
on the right in the $M_2$ versus $\mu$ plane. 
The yellow, dark green, light green and grey points have $\lsp$ masses of 15--25~GeV, 25--35~GeV, 
35--50~GeV and 50--60~GeV, respectively. Points which might be excluded either due to the factor 2 uncertainty 
in the implementation of the SMS limit for the $\tau$-dominated case from  the CMS analysis \cite{CMS-PAS-SUS-12-022} or by the ATLAS $2\tau$'s + $E_T^{\rm miss}$ analysis~\cite{ATLAS-CONF-2013-028} are 
flagged as triangles in a lighter color shade.
\label{fig:muM2} }
\end{figure}

The cross section for neutralino scattering on nucleons is dominated by the Higgs exchange diagram hence is driven by the higgsino fraction. For neutralinos below 30 GeV the cross section is mostly within one order of magnitude of the current XENON100 limit. It can however be much suppressed when the LSP has a small higgsino fraction. This occurs when the neutralino mass is near $m_Z/2$ or $m_h/2$ or when the light neutralino is purely bino and accompanied by light staus and light selectrons/smuons.
 
The interplay with indirect DM detection is also interesting. Figure~\ref{fig:indirect} shows $\sigma v$ corresponding to DM annihilation in the galaxy in either the $b\bar{b}$ or $\tau\tau$ channel. 
The  upper limit  from Fermi-LAT indirect searches for photons produced from DM annihilation in dwarf spheroidal galaxies allows  to constrain  a small subset of the points with light  DM annihilating into $\tau\tau$.   Some of these points are also in the region probed by  Fermi-LAT searches
for DM annihilation in subhalos  ~\cite{Berlin:2013dva} or from the Galactic Center~\cite{Hooper:2011ti}, the latter
bounds however depend  on the assumed DM profile. However, a large fraction of allowed points corresponding to  $\mneut>30 ~{\rm GeV}$ are several orders of magnitude below the current limits whether their main annihilation channel be into $\tau\tau$ or $b\bar{b}$. 
(For completeness we note that $\sigma v$ goes down to $\approx 10^{-34}~{\rm cm^2s^{-1}}$ in the $\tau\tau$ channel and down to $\approx10^{-31}~{\rm cm^2s^{-1}}$ in the $b\bar{b}$ channel.)

 \begin{figure}[t!]\centering
\includegraphics[width=0.48\textwidth]{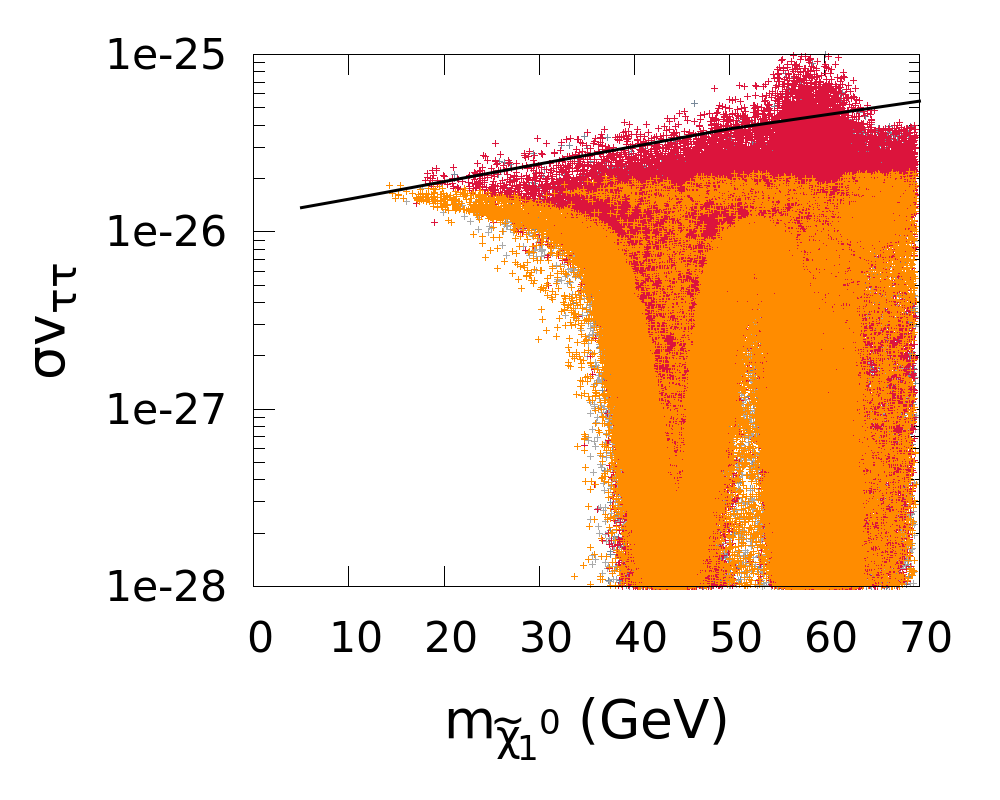}
\includegraphics[width=0.48\textwidth]{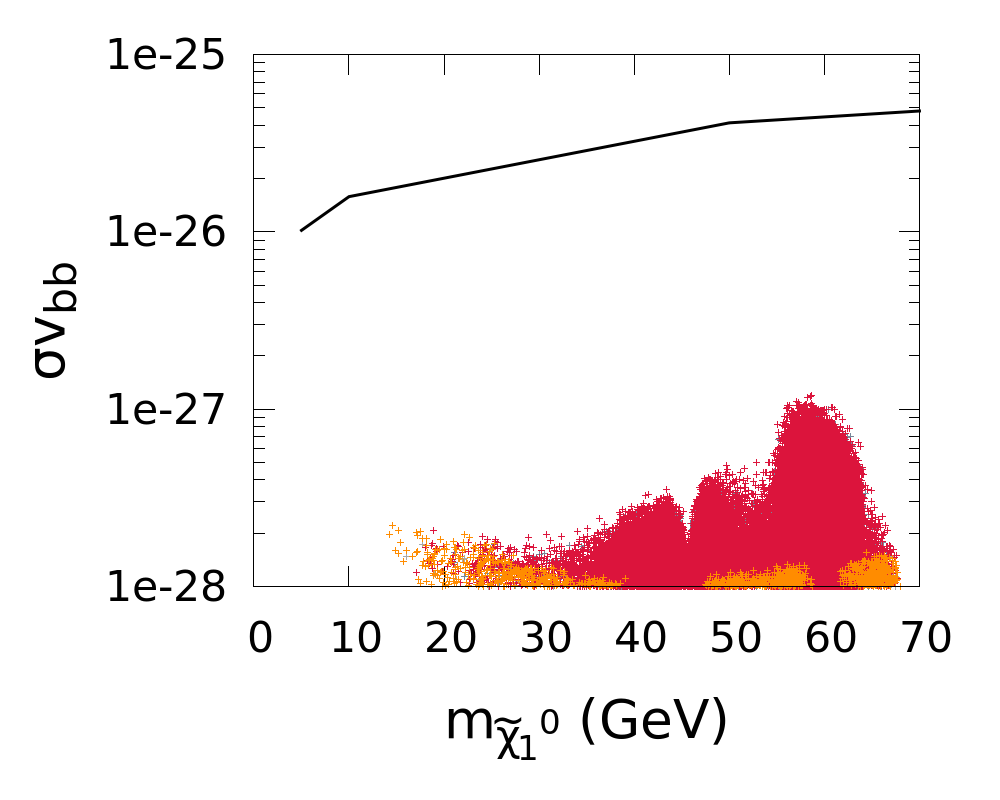}
\vspace*{-2mm}
\caption{Cross sections $\sigma v$ (in ${\rm cm^2s^{-1}}$) for indirect DM detection in the $\tau\tau$ (left) and $b\bar b$ (right) channels.  The black line shows the 95\% CL Fermi-LAT bound. 
Only points which satisfy the relic density and direct detection constraints are shown; 
following the color code of Fig.~\ref{fig:relic}, red (orange) points have $\Omega h^2<0.131$ ($0.107<\Omega h^2<0.131$).
\label{fig:indirect} }
\end{figure}

We next consider the implications for Higgs signal strengths $\mu$ relative to SM expectations in various channels. 
There are two features that can lead to modifications of the signal strengths in our scenario: a light neutralino 
and a light stau. The presence of a light neutralino can lead to a sizeable invisible decay width, thus leading 
to reduced signal strengths in all channels.
A  light stau  can contribute to the loop-induced $h\gamma\gamma$ coupling~\cite{Carena:2012gp}. In particular, heavily mixed staus can lead to enhanced signal strengths in the di-photon channels, while not affecting the signal in other decay channels. This is illustrated in Fig.~\ref{fig:mugamgam} (top panels). Here, the points with an enhanced $\mu(gg,\gamma\gamma)\equiv \mu(gg\to h\to \gamma\gamma)$\footnote{We use the notation $\mu(X,Y)\equiv \mu(X\to h\to Y)$.} are the ones with light, maximally mixed staus; these points occur only for $m_{\lsp}>25$~GeV and their signal strengths in the $VV$ ($WW$ or $ZZ$) and $b\bar{b}/\tau\tau$ channels do not differ significantly from 1, as can be seen in the bottom panels of Fig.~\ref{fig:mugamgam}. 
To achieve large stau mixing, we need $\mu\gtrsim 400$~GeV, so $\charg$ and $\tilde\chi^0_2$ are heavy in this case. 
Moreover, the scenarios with mixed staus require light selectrons/smuons in order to achieve low enough $\Omega h^2$.  Therefore these points are mostly constrained by the ATLAS results from direct slepton searches. 

\begin{figure}[t!]\centering
\includegraphics[width=0.48\textwidth]{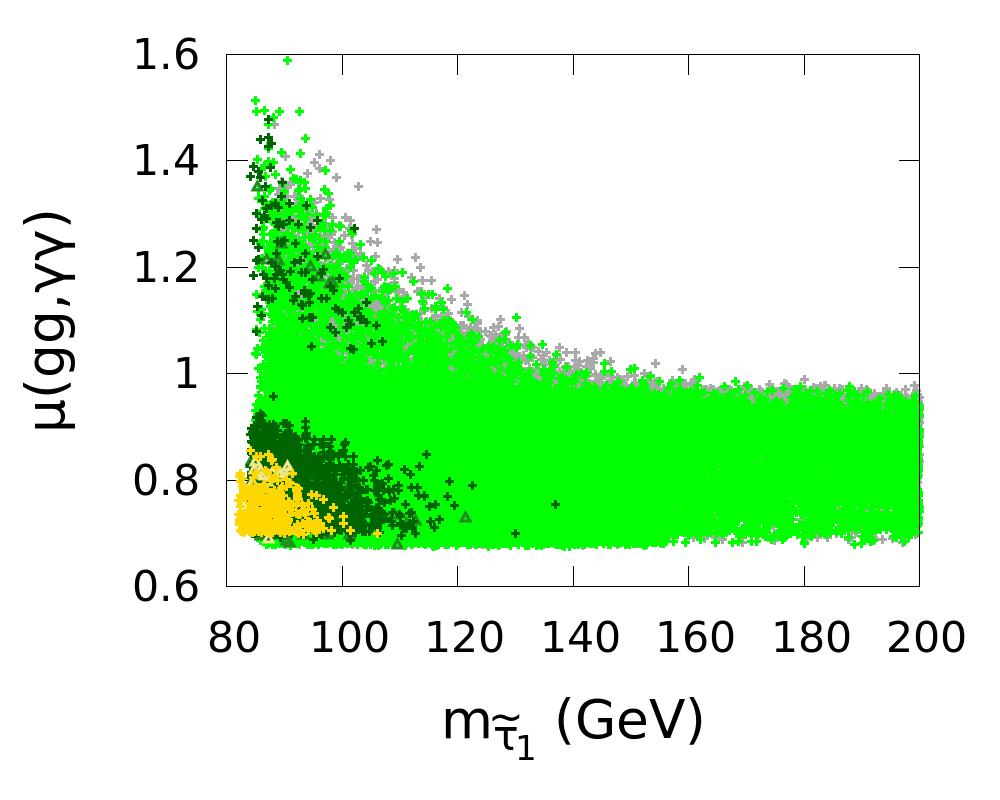}
\includegraphics[width=0.48\textwidth]{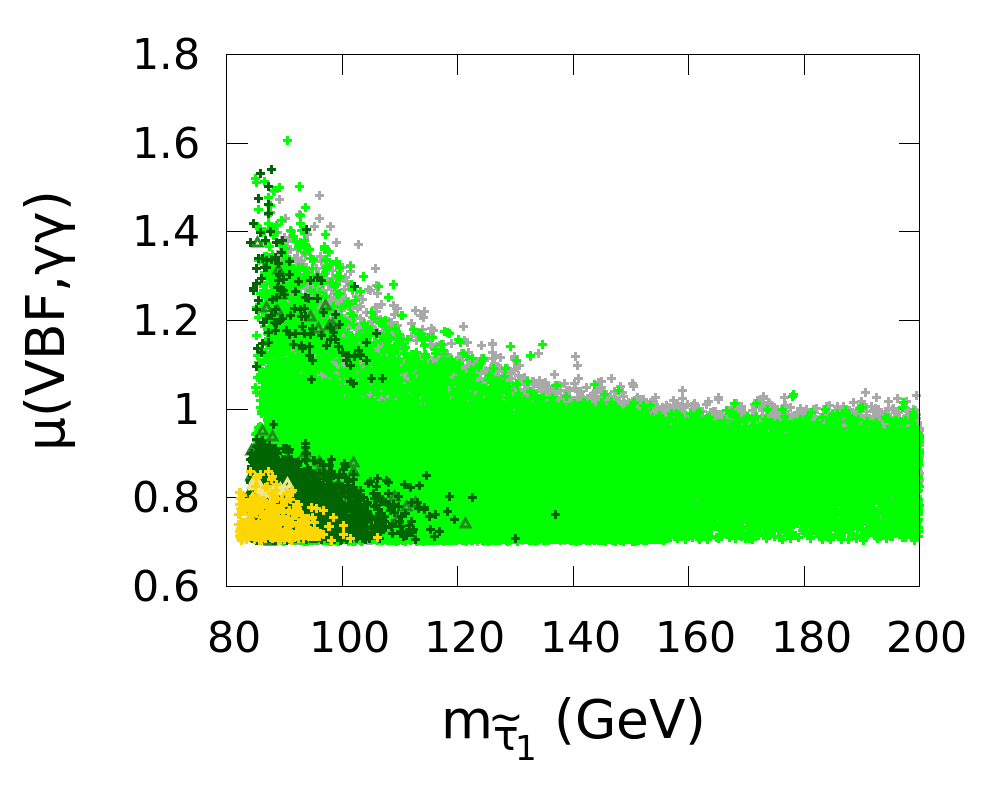} \\
\includegraphics[width=0.48\textwidth]{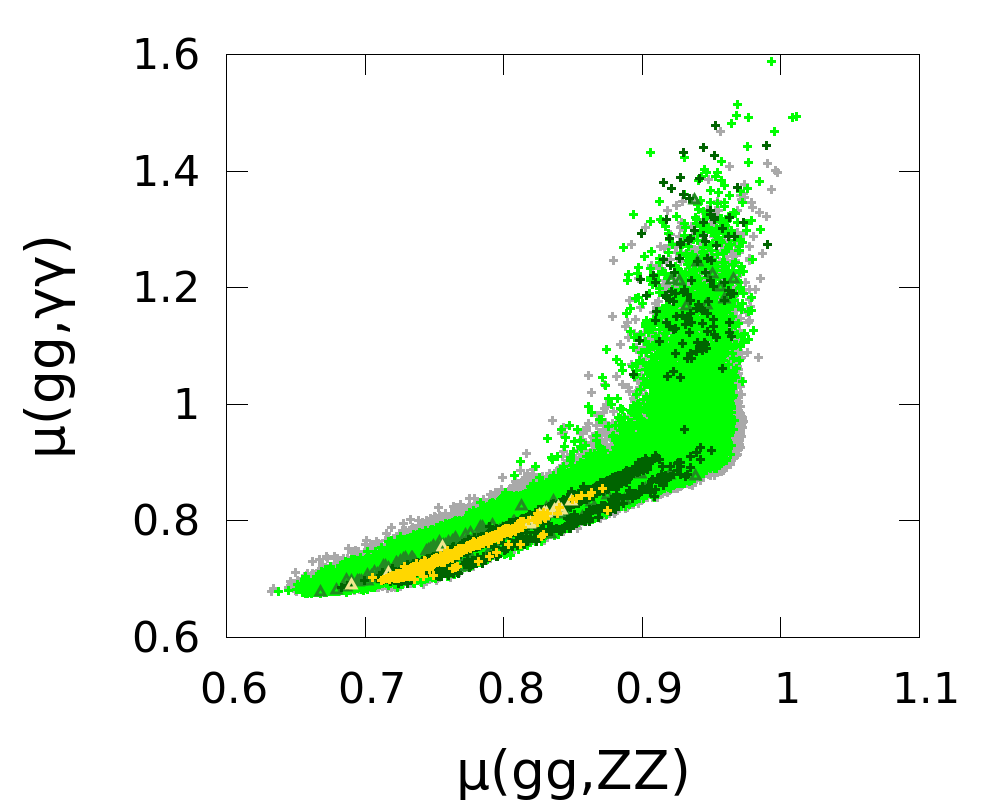}
\includegraphics[width=0.48\textwidth]{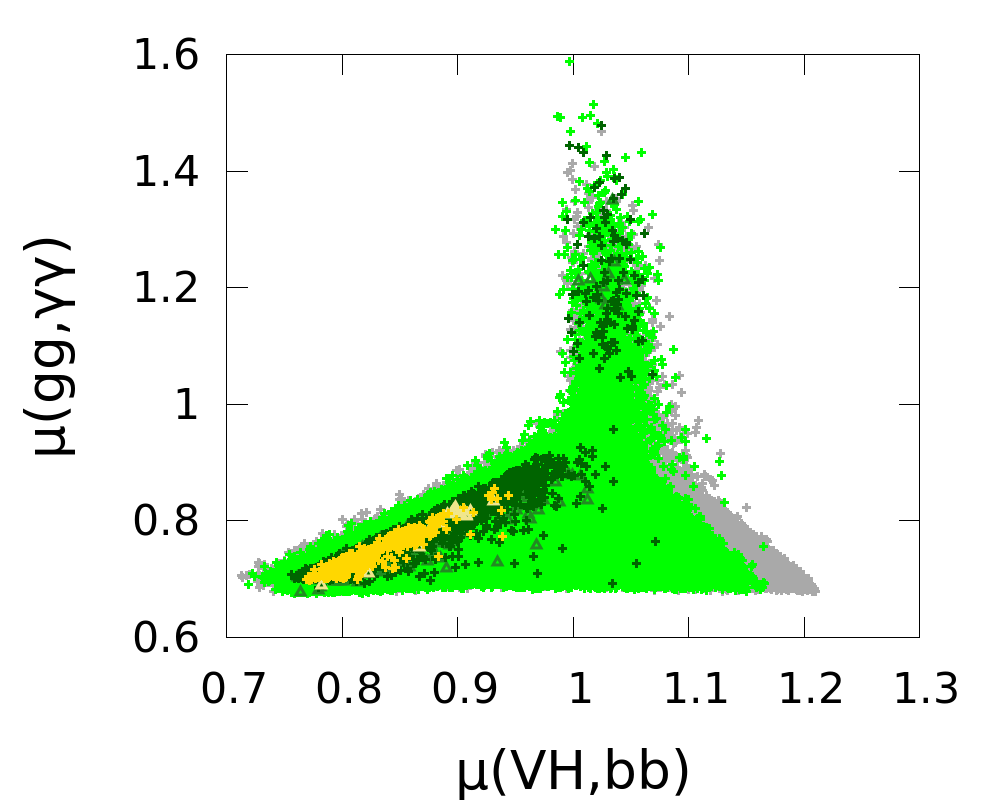} \\
\caption{Implications of the light neutralino dark matter scenario for Higgs signal strengths. 
Same color code as in Fig.~\ref{fig:muM2}.
\label{fig:mugamgam} }
\end{figure}

\begin{figure}[t!]\centering
\includegraphics[width=0.48\textwidth]{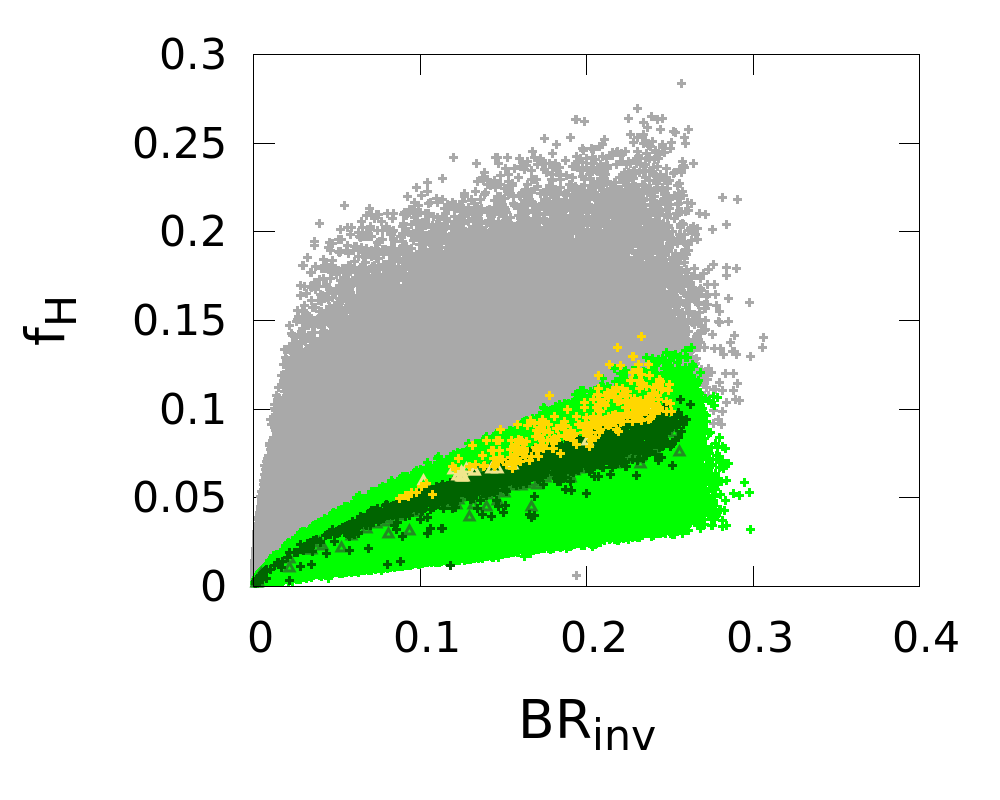} 
\includegraphics[width=0.48\textwidth]{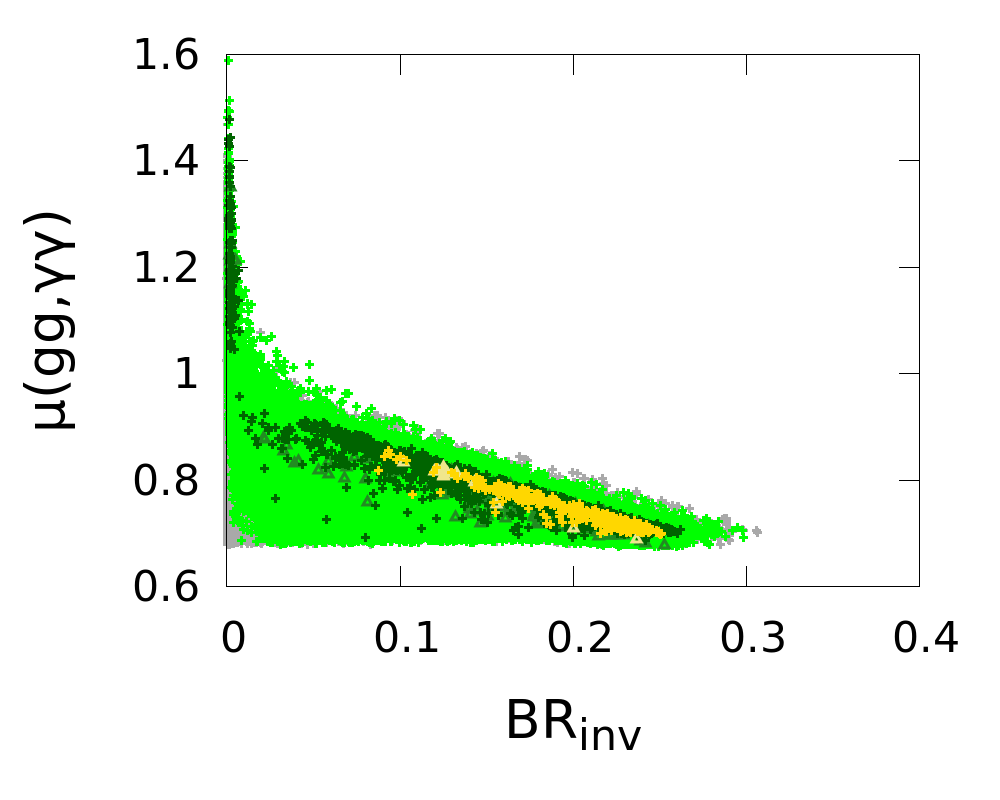}
\caption{Implications of the light neutralino dark matter scenario for invisible $h$ decays. 
Same color code as in Fig.~\ref{fig:muM2}.
\label{fig:inv} }
\end{figure}

The bulk of the light $\lsp$ points
however features a  reduced $\mu({gg,\gamma\gamma})\approx 0.7-0.9$.  This occurs when the stau  has only a mild effect on  $h\gamma\gamma$ and the invisible decay is sizeable, see Fig.~\ref{fig:inv}. In particular, for the very light neutralino sample with $m_{\lsp}=15-25$~GeV the light  $\tilde\tau_R$ needed for DM constraints does not help in increasing the 
$h\gamma\gamma$  coupling, hence all these points (in yellow  in Fig.~\ref{fig:mugamgam})
have a reduced signal strength.  Note also that for the points with $\mu({gg,\gamma\gamma})<1$, also  $\mu({gg,VV})$ is suppressed, see bottom-left panel in Fig.~\ref{fig:mugamgam}. Here, suppression of the gluon-fusion process by the stop-loop contribution also plays a role on top of the effect from invisible decays of the Higgs boson. 
Associated $VH$ production on the other hand is unaffected by this, and since the Higgs branching ratio into $b\bar{b}$ can be enhanced or suppressed $\mu({VH,b\bar{b}})$ can be above or below 1, as can be seen in the bottom-right panel of Fig.~\ref{fig:mugamgam}.

The invisible branching ratio of the Higgs can vary up to $\approx 30\%$ (the maximum allowed by the global Higgs fit) and is large for a large higgsino fraction of the LSP modulo kinematic factors, as illustrated in Fig.~\ref{fig:inv}. For this reason, the invisible width can be large for neutralinos below 35~GeV, leading to suppressed Higgs signals in all channels. Moreover, the points  with $\mu(gg,\gamma\gamma)>1$ have a small invisible width since they correspond to mixed staus and a small higgsino fraction (because large stau mixing requires large $\mu$) as mentioned above. The invisible width is also suppressed for $\mneut\approx m_Z/2$ because of the small higgsino fraction as well as near $m_h/2$ for kinematical reasons.  
It should be noted, however, that the determination of the allowed values of the branching ratios of the Higgs into 
invisible channels do, to some extent, depend on the way in which theoretical uncertainties on the Higgs 
cross sections are included in the analysis~\cite{Djouadi:2013qya}.

Future experimental results on the various Higgs signals will help constraining MSSM scenarios with a light neutralino, as can be expected from the 14~TeV projections of ATLAS~\cite{ATL-PHYS-PUB-2012-004} and CMS~\cite{CMS-NOTE-2012-006} for ${\cal L} = 300\fbi$. The estimated precision on the signal strengths is of the order of $10$\% in several channels of interest, including $h \to \gamma\gamma$ and $h \to ZZ$. As can be seen in Fig.~\ref{fig:mugamgam}, this will help discriminating between the various scenarios---in particular, the points with $m_{\lsp}=15$--$25$~GeV have $\mu^{\max}(gg,\gamma\gamma) \approx \mu^{\max}(gg,ZZ) \approx 0.86$. A better determination of the invisible decays of the Higgs boson should also probe further the remaining parameter space, both from a global fit to the properties of the Higgs and from direct searches for Higgs decaying invisibly. In the latter case, the projected upper bound is found to be $\brinv \lesssim 0.17$ at 95\% CL for 100$\fbi$\ at 14~TeV~\cite{Ghosh:2012ep}.

\section{Conclusions}

We find that although the most recent LHC limits on Higgs properties and on direct production of SUSY particles impose strong constraints on the model, neutralino LSPs as light as 15 GeV can be compatible with all data. These scenarios require light staus below about 100~GeV and light charginos below about 200~GeV. They will be further probed at the LHC at 13--14~TeV through searches for sleptons and electroweak-inos, as well as  through  a more precise determination of the Higgs couplings. Moreover, improving  the spin-independent direct detection limits  by an order of magnitude  will cover the whole range of $\lsp$ masses below $\approx 35$~GeV.

\subsection*{Note added}

While this Letter was in preparation, Ref.~\cite{Calibbi:2013poa} appeared with very similar 
conclusions as the ones reached here.  While Ref.~\cite{Calibbi:2013poa} features a  better implementation 
of the limits from multi-tau plus $E_T^{\rm miss}$ searches, based on a reinterpretation of the ATLAS results~\cite{ATLAS-CONF-2013-028}, it does not include or discuss implications for the Higgs sector. 
Another difference is that we consider a wider range of MSSM scenarios; in particular we include 
the possibility of light selectrons/smuons  as well as of light winos.


Moreover, just prior to publication, another paper~\cite{Arbey:2013aba} discussing the possibility of light neutralinos   appeared. This work is complementary to ours as it concentrates on light neutralinos  almost degenerate with light sbottoms  which we do not consider.

\section*{Acknowledgements} 

We thank U. Laa, A. Lessa, D. Proschofsky-Spindler and W. Waltenberger for 
helpful discussions regarding Simplified Models Spectra and for the possibility to use the {\tt SmodelS} 
approach for this work.
Moreover, we thank T.~Han and A.~Pukhov for useful discussions, 
M.~ Gauthier-Lafaye for his help in using the MUST-cluster in Annecy, 
and the reactor physics group of LPSC Grenoble for the use of part of their computer resources. 

This study originated from the workshop ``Implications of the 125~GeV Higgs boson'', 
18--22 March at LPSC Grenoble. Part of the work was performed during the 
``Physics at TeV Colliders'', 12--21 June 2013 in Les Houches. 
RG wishes to thank LPSC Grenoble for hospitality and the Department of Science and Technology, 
Government of India, for support under grant no. SR/S2/JCB-64/2007. 
GB and SK thank the Aspen Center for Physics for hospitality during the final stage of this work. 
Partial funding by the French ANR~DMAstroLHC is gratefully acknowledged.


\providecommand{\href}[2]{#2}\begingroup\raggedright\endgroup

\end{document}